\documentclass[11pt,twoside]{article}  % Leave intact
\usepackage{apn3conf}

\begin{document}   % Leave intact

%-----------------------------------------------------------------------
%		            Paper Title 
%-----------------------------------------------------------------------
% Enter the title of the paper.
%
% EXAMPLE: \title{A Breakthrough in Astronomical Software Development}
% 
% If your title is so long as to fill the page header when you print it,
% then please supply a short form as a \titlemark.
%
% EXAMPLE: 
%  \title{Rapid Development for Distributed Computing, with Implications
%         for the Virtual Observatory}
%  \titlemark{Rapid Development for Distributed Computing}
%

\title{Two New Close Binary Central Stars of Planetary Nebulae from a Critically Selected Southern Hemisphere Sample}
\titlemark{Two New Close Binary CSPNe}

%-----------------------------------------------------------------------
%		          Authors of Paper
%-----------------------------------------------------------------------
% Enter the authors followed by their affiliations.  The \author and
% \affil commands may appear multiple times as necessary (see example
% below).  List each author by giving the first name or initials first
% followed by the last name.  Authors with the same affiliations
% should grouped together. 
%
% EXAMPLE: \author{Margaret Meixner\altaffilmark{1}, Letizia Stanghellini,
%			Howard Bond} 
%          \affil{Space Telescope Science Institute, 
%                 3700 San Martin Dr.,  Baltimore, MD 21218}
%          \author{Joel Kastner}
%          \affil{Rochester Institute of Technology}
%
%          \altaffiltext{1}{Astronomy Department, UIUC}
%
% In this example, the first three authors, "Meixner", "Stanghellini", and
% "Bond" are affiliated with "STScI".  "Meixner" has an alternate 
% affiliation with the "Astronomy Department at UIUC".  The fourth author,
% "Kastner", is affiliated with "Rochester Institute of Technology"

\author{Todd Hillwig}
\affil{Department of Physics and Astronomy, Georgia State University}

%-----------------------------------------------------------------------
%			 Contact Information
%-----------------------------------------------------------------------
% This information will not appear in the paper but will be used by
% the editors in case you need to be contacted concerning your
% submission.  Enter your name as the contact along with your email
% address.
% 
% EXAMPLE:  \contact{Dennis Crabtree}
%           \email{crabtree@cfht.hawaii.edu}
%

\contact{Todd Hillwig}
\email{thillwig@chara.gsu.edu}

%-----------------------------------------------------------------------
%		      Author Index Specification
%-----------------------------------------------------------------------
% Specify how each author name should appear in the author index.  The 
% \paindex{ } should be used to indicate the primary author, and the
% \aindex for all other co-authors.  You MUST use the following
% syntax: 
%
% SYNTAX:  \aindex{LASTNAME, F. M.}
% 
% where F is the first initial and M is the second initial (if
% used).  This guarantees that authors that appear in multiple papers
% will appear only once in the author index.  
%
% EXAMPLE: \paindex{Crabtree, D.}
%          \aindex{Manset, N.}        
%          \aindex{Veillet, C.}        
%
% NOTE: this information is also used to build the author list that
% appears in the table of contents.  Authors will be listed in the order
% of the \paindex and \aindex commmands.
%

\paindex{Hillwig, T. C.}
%\aindex{ }      Remove this line if there is only one author

%-----------------------------------------------------------------------
%		      Author list for page header	
%-----------------------------------------------------------------------
% Please supply a list of author last names for the page header. in
% one of these formats:
%
% EXAMPLES:
% \authormark{LASTNAME}
% \authormark{LASTNAME1 \& LASTNAME2}
% \authormark{LASTNAME1, LASTNAME2, ... \& LASTNAMEn}
% \authormark{LASTNAME et al.}
%
% Use the "et al." form in the case of seven or more authors, or if
% the preferred form is too long to fit in the header.

\authormark{Hillwig}

%-----------------------------------------------------------------------
%			Subject Index keywords
%-----------------------------------------------------------------------
% Enter up to 6 keywords describing your paper.  These will NOT be
% printed as part of your paper; however, they will be used to
% generate an object index and a subject index for the proceedings.  
% There is no standard list,  however, individual object names are
% encouraged and one or two word descriptions of the topics (e.g.MHD, 
% ionized gas) are useful. 
%
% EXAMPLE:  \keywords{NGC 7027, AFGL 2688, HD 161796, binary stars,
%                      dust,  molecular gas}
%

\keywords{NGC 6026, NGC 6337, binary stars, photometry}

%-----------------------------------------------------------------------
%			       Abstract
%-----------------------------------------------------------------------
% Type abstract in the space below.  Consult the User Guide and Latex
% Information file for a list of supported macros (e.g. for typesetting 
% special symbols). Do not leave a blank line between \begin{abstract} 
% and the start of your text.

\begin{abstract}          % Leave intact
I  present the results of time-resolved photometry of a selection of
central stars of planetary nebulae.  The study reveals periodic variability
in two of the eight central stars observed, those of NGC 6026 and NGC 6337.
The variability matches that expected from a binary system in which a hot
primary irradiates a cooler secondary star.

\end{abstract}

%-----------------------------------------------------------------------
%			      Main Body
%-----------------------------------------------------------------------
% Place the text for the main body of the paper here.  You should use
% the \section command to label the various sections; use of
% \subsection is optional.  Significant words in section titles should
% be capitalized.  Sections and subsections will be numbered
% automatically. 
%
% EXAMPLE:  \section{Introduction}
%           ...
%           \subsection{Our View of the World}
%           ...
%           \section{A New Approach}
%
% It is recommended that you look at the sample papers, sample1.tex
% and sample2.tex, for examples for formatting references, footnotes,
% figures, equations, html links, lists, and other special features.  

\section{Introduction}

Soker (1997) explained planetary nebula (PN) structures using orbital
interactions in a binary system, with the companion to the PN progenitor 
being either stellar or substellar (brown dwarf or planet).  Using this 
theory he classified a large number of PNe, based on their morphology, as 
either: single progenitor, close stellar companion with no common 
envelope (CE) phase, close stellar companion with a CE phase, or 
substellar companion with a CE phase.  These predictions provide a good 
basis for testing the binary theory of PN shaping.

It is understood that for close stellar companions undergoing a CE phase, a 
significant fraction should become systems with orbital periods on the 
order of a few days or less.  To 
date, thirteen central stars of planetary nebulae (CSPNe) have been 
identified as close binaries, with orbital periods determined to be $< 16$ 
days (Bond 2000).  Bond (2000) gives the fraction of 
detectable close binaries from a random sample as $\sim 10-15\%$.

Here I present the use of time 
resolved photometry of CSPNe to search for the sinusoidal variations in 
brightness that would be associated with an irradiated hemisphere of a 
stellar companion to the central star.  Eight southern hemisphere CSPNe
classified by Soker (1997) as having a stellar companion that underwent
a CE phase were observed.  The results support the 
classification of NGC 6026 and NGC 6337 as close binary stars.  I 
present orbital periods for these systems.  My detected binary fraction 
is also compared to that discovered previously.

\section {Observations}

The photometric data consist of V-band CCD photometry obtained from 30
April - 4 May, 2002 at the Cerro Tololo Interamerican Observatory 
(CTIO) 0.9m telescope with the T2K Imager.  The observations were 
carried out such that every object was observed each night for at least 
one hour and the five-night spacing was altered to reduce as many aliases 
as possible in the period search algorithm to be utilized.

The exposures were reduced using the DAOPHOT package of IRAF.
The resulting photometry was then analyzed by 
incomplete ensemble photometry (Honeycutt 1992).  The zero points are 
instrumental magnitudes dependent on the instrumental response of the 
CTIO 0.9m system.  

The light curves were analyzed with {\it periodogram}, a 
program which uses the period search technique of Scargle (1982) as 
modified by Horne \& Baliunas (1986).

Six of the eight stars showed no unambiguous variability.  Five of the
six: He 2-141, NGC 3132, NGC 4361, NS 238, and Sp 3, where observed
on at least four nights, with $55\leq t \leq 78$ minutes of consecutive
observations on at least one of those nights.  K 1-1 was observed on
three nights with a maximum duration of 64 minutes.

\section {NGC 6026}

\begin{figure}[t]
\begin{center}
\epsscale{.50}
\plottwo{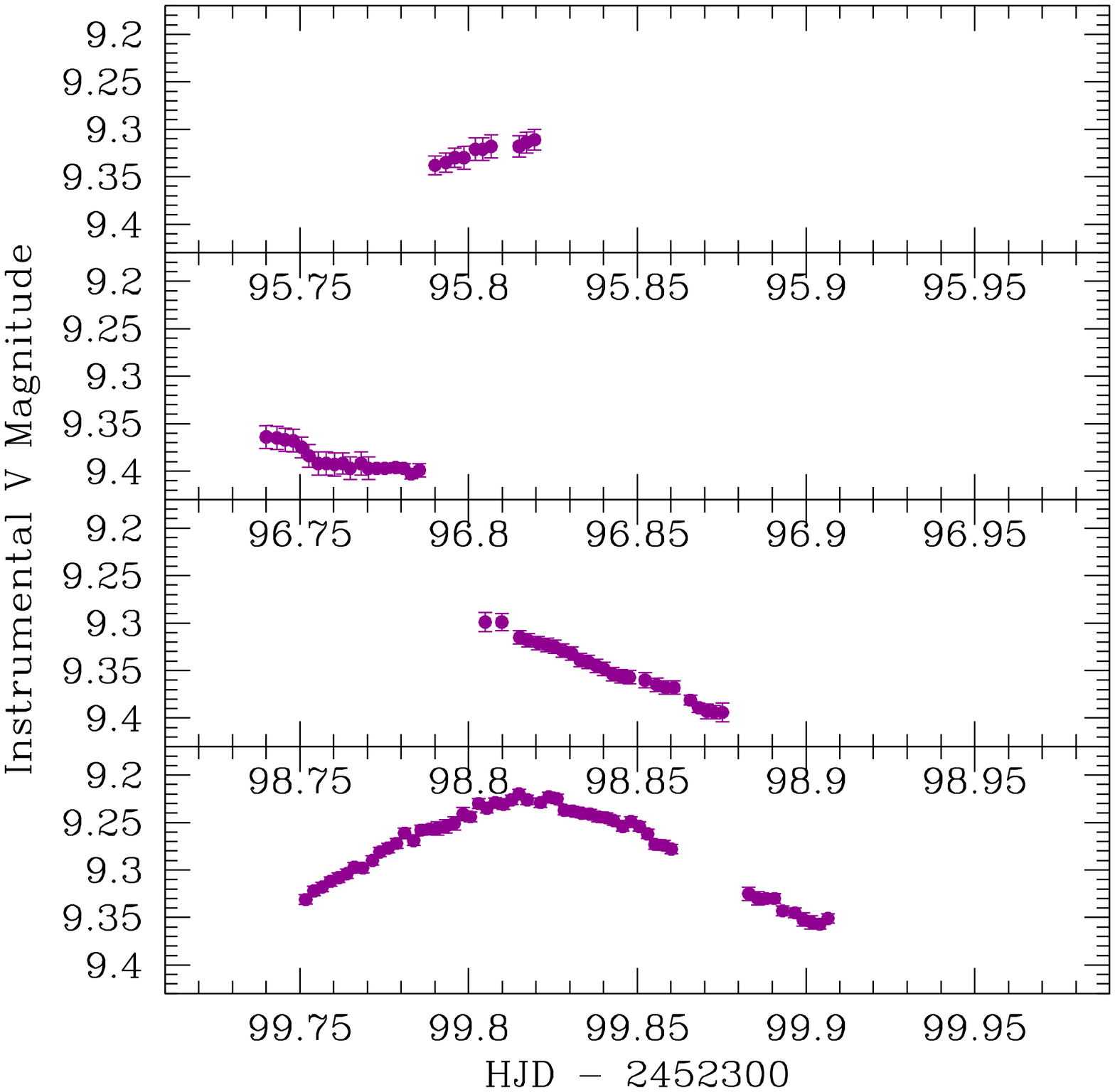}{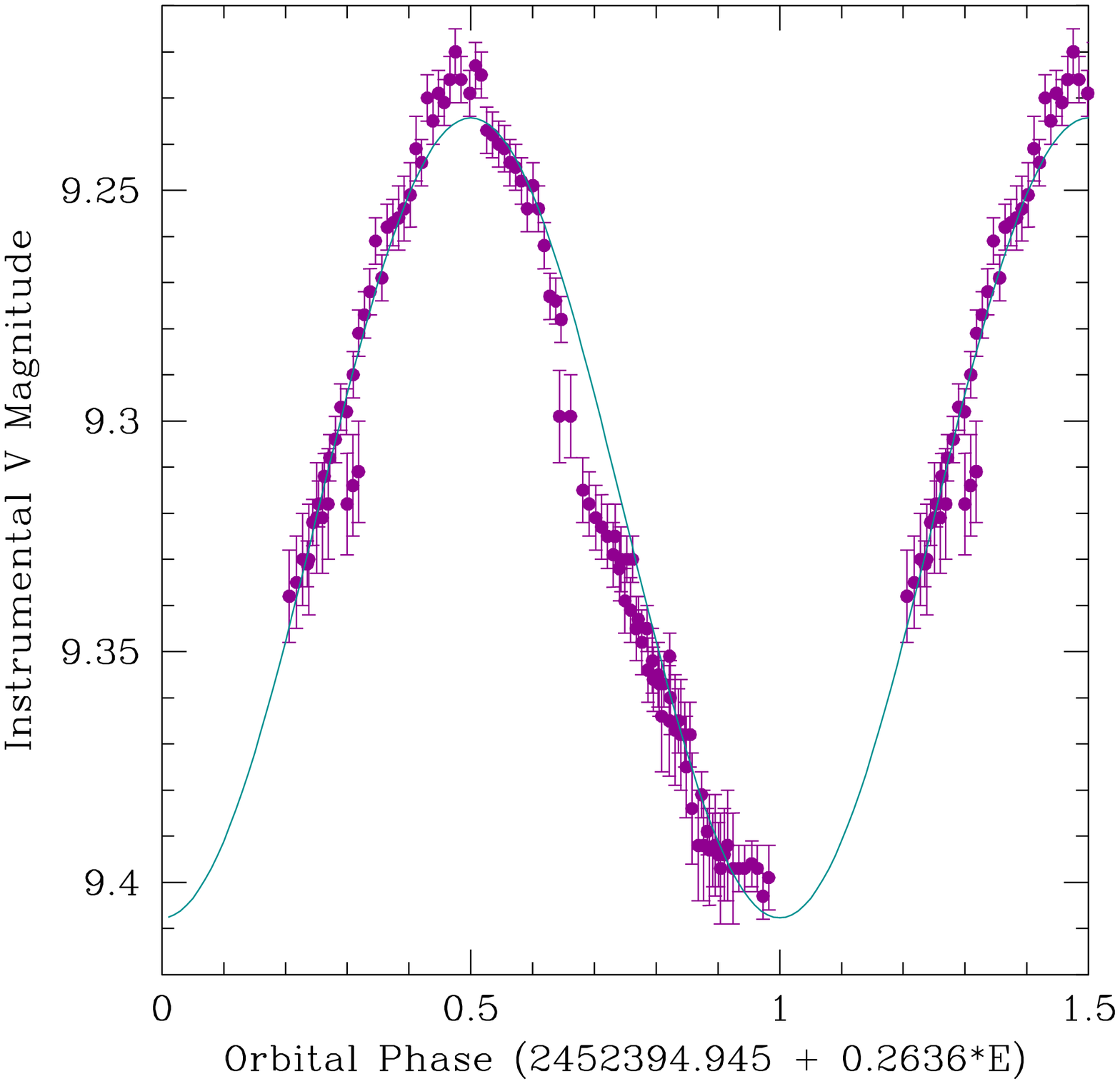}
\end{center}
\caption{Light curve for NGC 6026 by night (\textbf{left}) and the
phase-folded light curve of NGC 6026
for the ephemeris in equation \ref{eq2}(\textbf{right}).}\label{6026}
\end{figure}

The light curve of NGC 6026 (Figure \ref{6026}) shows obvious
variability and suggests a well-behaved periodic nature over the four
nights on which observations were made.
The periodogram for NGC 6026 showed periodic
variability with a strong one day alias, resulting in a preliminary period
of 0.263 days.  This 
period was taken as a starting point for fitting a sine curve of the form
\begin{equation}
V=V_{mean}+Ksin\Biggl[\frac{2\pi(T-T_o)}{P}+\phi\Biggr]
\end{equation}
where $V$ is the instrumental magnitude, $K$ is the semiamplitude,
and $P$ is the period; either $T_o$ or $\phi$ is a fitted parameter.
For the photometry we fit for $T_o$ plus the remaining parameters and
set $\phi$ = 0.75 as is appropriate for variation dominated by an
irradiation effect.  This produces an ephemeris 
with minimum light at phase zero:
\begin{equation}
T=2452394.945(1)+0.2636(2)E.
\label{eq2}
\end{equation}
Figure \ref{6026} shows the light curve folded on this ephemeris and the
sine curve fit.  The rms variation of the data from the fitted sine curve 
is 0.010 magnitude.

The remaining parameters from the sine fitting are 
$V_{mean}=9.3210(7)$ mag (instrumental) and $K=0.087(1)$ mag.

\section {NGC 6337}

\begin{figure}[t]
\begin{center}
\epsscale{.50}
\plottwo{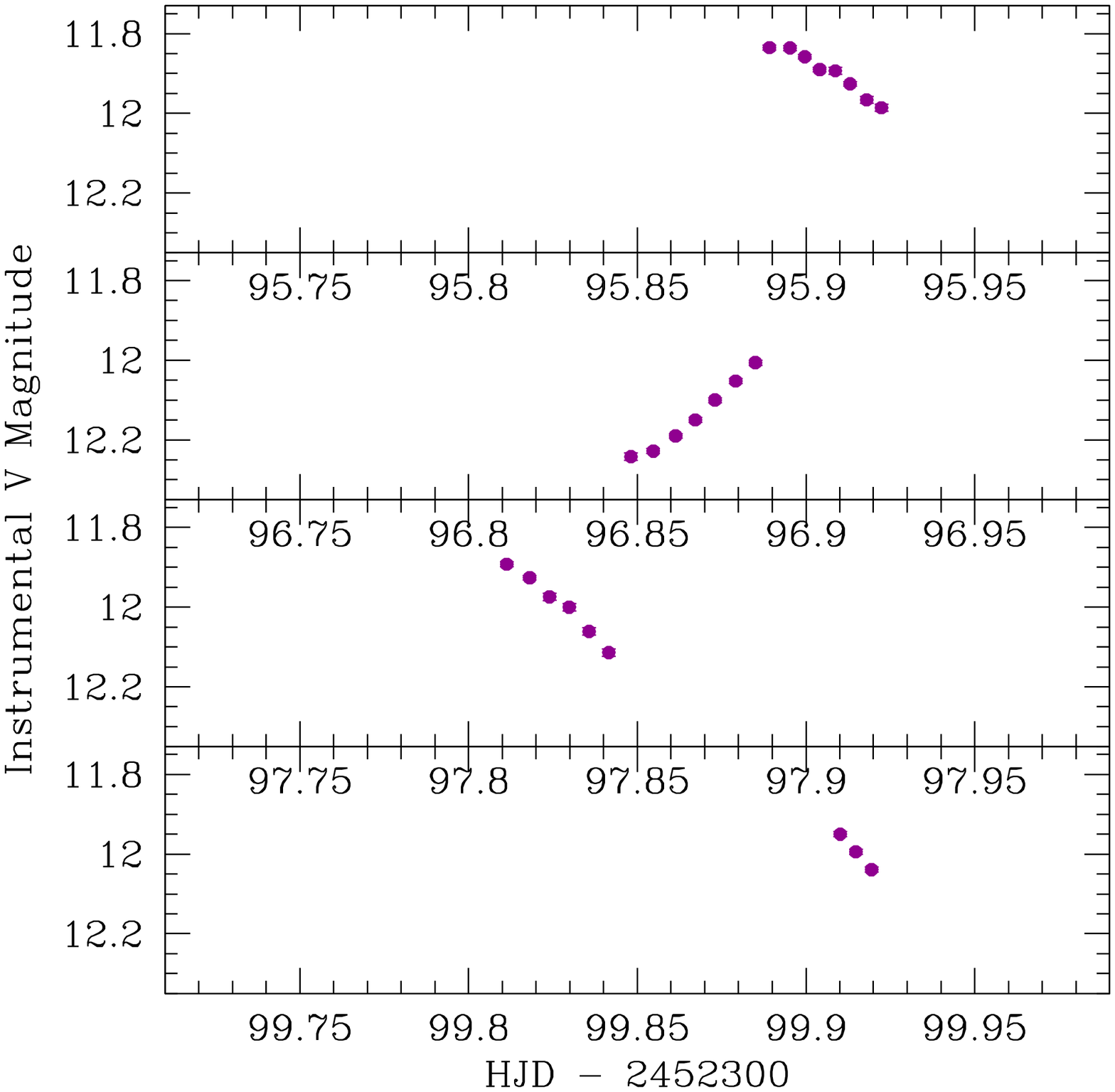}{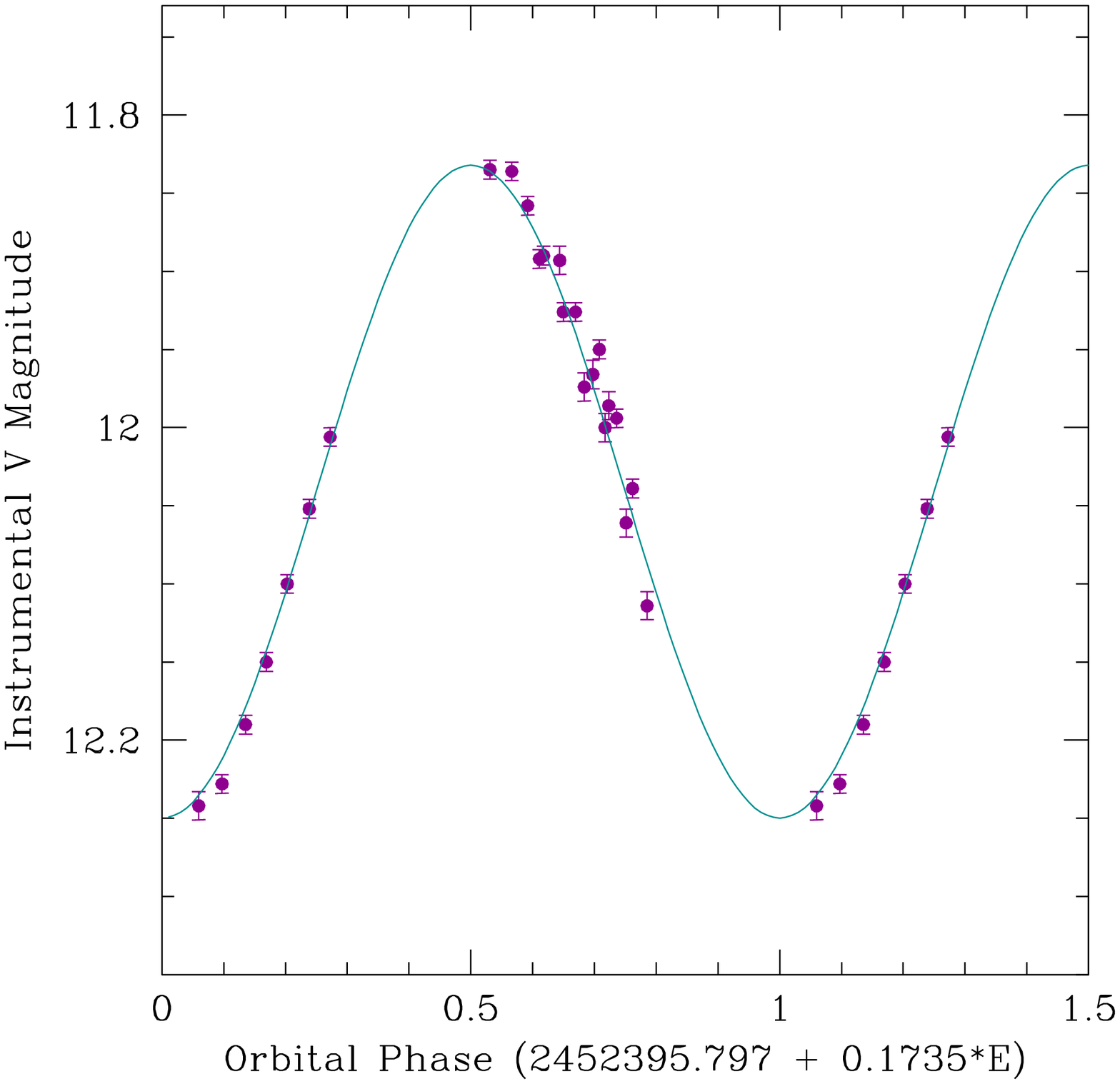}
\end{center}
\caption{Light curve for NGC 6337 by night (\textbf{left}) and the
phase-folded light curve of NGC 6337
for the ephemeris in equation \ref{eq3}(\textbf{right}).}\label{6337}
\end{figure}

As with NGC 6026, the light curve of NGC 6337 in Figure \ref{6337} shows
apparently well-behaved periodic variability.
The periodogram showed periodic variability with a strong one day alias.
The peak with the most power was located at a period of 0.173 days.
Since NGC 6337 appears as a ring, the nebular background close to the 
central star is at a minimum, allowing more accurate photometry.  This, 
along with the obvious slope to each night of observation allowed all 
but the highest peak to be rejected.  The corresponding period was 
taken as a starting point for fitting a sine curve of the same form 
as Equation 1, for NGC 6026.  The derived photometric ephemeris for 
minimum light in NGC 6337 is then
\begin{equation}
T=2452395.7969(5)+0.1735(3)E.
\label{eq3}
\end{equation}
Figure \ref{6337} shows the light curve folded on this ephemeris and
the sine curve fit.  The rms variation of the data from the fitted
sine curve is 0.016 magnitude.

The remaining parameters from the sine fitting are 
$V_{mean}=12.041(1)$ mag (instrumental) and $K=0.209(3)$ mag.

\section {Comments on Masses, Inclinations, and Binary Fractions}

The short orbital periods allow limits to be placed on several of
the binary system parameters. Taking the masses of both 
CSs to be $\sim 0.6 M_\odot$, typical of CSPNe, Kepler's 2nd law gives
the  binary separation as a function of secondary mass.  The small
deviation from a sine curve, short orbital
periods, and significant reflection effects observed
in the two systems suggest that the companions are cool dwarf stars
which do not fill, or just fill, their Roche lobes.
To find an upper limit for the secondary masses, I will assume that
both secondaries are Main Sequence stars and do not fill their
Roche lobes.  This condition is met for the two systems,
according to their corresponding orbital periods and for
$M_1 = 0.6 M_\odot$, by $M_2(NGC 6026) \leq 0.8 M_\odot$ and
$M_2(NGC 6337) \leq 0.3 M_\odot$.  These correspond to binary
separations of $a(NGC 6026) \leq 1.93 R_\odot$ and
$a(NGC 6337) \leq 1.26 R_\odot$.

A few comments may be made about the binary system inclinations as well.
Corradi et al. (2000) show that the ring-like structure of NGC 6337 is
the narrow waist of a nearly pole-on bipolar PN.  The inclination of 
the PN then must be $\leq 15^\circ$ for the bright inner ring to have no 
apparent ellipticity.
NGC 6026 does not appear ring-shaped, but as a partial ellipse with a 
very faint south-east edge.  If this PN is what Bond (2000) refers to as
a ``wedding ring'' PN, then the inclination must be intermediate and can be
roughly determined from the ellipticity of the observed PN.  The ellipse
measures $\sim32\arcsec \times 42\arcsec$, giving an inclination of
$\sim40^\circ$.  If the binary mechanism for shaping PNe is 
correct, then the close binary CSs must have inclinations equal to
those of the PNe.

Finally, previous studies have found the fraction of detectable close
binary CSPNe to be $\sim10-15\%$ (Bond 2000).  The binary fraction
found here is 25\%, 
much higher than previous studies, though the low number of statistics 
means that the results are reasonably similar.  Since the sample 
observed in this study was not randomly selected, but was based on 
the classifications of Soker (1997), it would be interesting to 
increase the number of CSPNe observed. If the 25\% binary fraction 
persisted, the binary mechanism for PN shaping would be strongly supported, 
specifically the scenario outlined by Soker (1997). 

\acknowledgements
I would like to thank Bill Bagnuolo, Doug Gies, and Bill Nelson for
their generous support and Todd Henry and Alberto Miranda for their 
assistance with the CTIO 0.9m telescope.  This research was funded
by Georgia State University.  

%-----------------------------------------------------------------------
%			      References
%-----------------------------------------------------------------------
% List your references below within the reference environment
% (i.e. between the \begin{references} and \end{references} tags).
% Each new reference should begin with a \reference command which sets
% up the proper indentation.  Observe the following order when listing
% bibliographical information for each reference:  author name(s),
% publication year, journal name, volume, and page number for
% articles.  Note that many journal names are available as macros; see
% the User Guide listing "macro-ized" journals.   
%
% EXAMPLE:  \reference Hagiwara, K., \& Zeppenfeld, D.\  1986, 
%                Nucl.Phys., 274, 1
%           \reference H\'enon, M.\  1961, Ann.d'Ap., 24, 369
%           \reference King, I.\ R.\  1966, \aj, 71, 276
%           \reference King, I.\ R.\  1975, in Dynamics of Stellar 
%                Systems, ed.\ A.\ Hayli (Dordrecht: Reidel), 99
% 
% Note the following tricks used in the example above:
%
%   o  \& is used to format an ampersand symbol (&).
%   o  \'e puts an accent agu over the letter e.  See the User Guide
%      and the sample files for details on formatting special
%      characters.  
%   o  "\ " after a period prevents LaTeX from interpreting the period 
%      as an end of a sentence.
%   o  \aj is a macro that expands to "Astron. J."  See the User Guide
%      for a full list of journal macros
%

% Do not place any material after the references section

\end{document}